\documentclass[jgrga,draft]{AGUTeX}
\usepackage{hyperref}
\usepackage{graphicx}

\usepackage{wrapfig}
\usepackage{caption}
\makeatletter
\def\ext@figure{}
\makeatother

\makeatletter
\def\ext@table{}
\makeatother

\setkeys{Gin}{draft=false}

\authorrunninghead{Winter {\it et al.}}
\titlerunninghead{{PROTON FLUX NOWCAST}}
\authoraddr{Corresponding author: Lisa Winter, Space Weather \& Effects Group, Atmospheric and Environmental Research, 131 Hartwell Avenue, Lexington, MA, USA. (lwinter@aer.com)}

\begin{document}

\newcommand{\ssr}{Space Science Reviews}
\newcommand{\planss}{Plan. S.S.}

\lefthead{GIC collaboration authors}
\righthead{Extreme Event GIC}

\title{Spectral Scaling Technique to Determine Extreme Carrington-level Geomagnetically Induced Currents Effects}  

\authors{Lisa M. Winter\altaffilmark{1}, Jennifer Gannon\altaffilmark{2}, Rick Pernak\altaffilmark{1}, Stu Huston\altaffilmark{1}, Richard Quinn\altaffilmark{1}, Edward Pope\altaffilmark{3}, Alexis Ruffenach\altaffilmark{4}, Pietro Bernardara\altaffilmark{4}, Nicholas Crocker\altaffilmark{4}}

\altaffiltext{1}{Atmospheric and Environmental Research, 131 Hartwell Avenue, Lexington, MA, USA.}
\altaffiltext{2}{Computational Physics Inc., 1650 38th St \# 105 W, Boulder, CO, USA, 80301.}
\altaffiltext{3}{Met Office, FitzRoy Road, Exeter, UK, EX1 3PB.}
\altaffiltext{4}{EDF Energy R\&D UK Centre Interchange, 81-85 Station Rd, Croydon, UK, CR02AJ.}

\begin{abstract}
Space weather events produce variations in the electric current in the Earth's magnetosphere and ionosphere. From these high altitude atmospheric regions, resulting geomagnetically induced currents (GICs) can lead to fluctuations in ground currents that affect the electric power grid and potentially overload transformers during extreme storms. The most extreme geomagnetic storm on record, known as the 1859 Carrington event, was so intense that ground-based magnetometers were saturated at high magnetic latitudes. The most reliable, un-saturated observation is the hour-resolution data from the Colaba Magnetic Observatory in India. However, higher frequency components -- fluctuations at second through minute time cadence -- to the magnetic field can play a significant role in GIC-related effects. We present a new method for scaling higher frequency observations to create a realistic Carrington-like event magnetic field model, using modern magnetometer observations. Using the magnetic field model and ground conductivity models, we produce an electric field model.
This method can be applied to create similar magnetic and electric field models for studies of GIC effects on power-grids.

\end{abstract}

\begin{article}
\section{Introduction}
The largest solar and geomagnetic storm on record was observed by Richard Carrington on 1 September 1859
\citep{2003JGRA..108.1268T}.  Carrington observed an abnormally large group of sunspots and subsequent solar flares in white light in his ground-based observatory.  This observed solar activity precipitated an abundance of activity - a coronal mass ejection, geomagnetic storm, aurorae at low geomagnetic latitudes - whose impacts could be detected at the surface of the Earth.  In modern space weather analyses, the Carrington Event has become an example of extreme geomagnetic storm conditions, and is often used as a worst case scenario in Geomagnetically Induced Current (GIC) planning. 

No storm as powerful as the Carrington Event has occurred since, but society is increasingly more reliant on electric power transmission than it was in 1859. Modern society relies upon the bulk power system for needs as diverse as preservation and distribution of food resources, public transportation, satellite communications, and information technology supporting e.g., emergency and medical services. With increased dependence, the infrastructures supporting the power grids are increasingly more complex and interconnected, making risk assessment, mitigation, and prevention more difficult. Additionally, with modern power grids operating closer to capacity than during the last extreme space weather event, the space weather effects from a future Carrington-level event are likely to be more wide-spread than the comparatively minor disruptions experienced in 1859 (e.g., communications disruptions to the telegraph systems). Such events can thus be considered high-impact (or high-consequence) and low-frequency.  Recent research in the field of Space Weather gives better understanding of how to mitigate, respond to, or even prevent geomagnetically induced current (GIC)-related damage on a global scale 
\citep{cite-NRCreport,2002JASTP..64..743E}, but it is also important to produce results for more localized regions \citep{2015GeoRL..42.6916N}.

In this article, we present a new technique to model and scale the magnetic and electric fields associated with space weather storms, which can be used to determine GIC-related damage to power grids at specified locations.
As an example of the technique, we present a model of an extreme space weather storm for the United Kingdom (UK). We create the expected magnetic field measurements for a scenario representative of a Carrington-level event and from here determine the expected electric field that drives GICs in long conductors. Magnetometer observations from twenty recent geomagnetic storms are used to determine a scaling factor between the strength of the storm and the power in the magnetic field observations. The magnetometer observations are described in Section~\ref{sect-observations} and the scaling technique is shown in Section~\ref{scaling}. The scaling factors are used to create a UK-specific Carrington-level storm magnetic field model in Section~\ref{sect-magfield}. The ground conductivity models and our electric field model are described in Sections~\ref{sect-conductivity} and \ref{sect-efield}, respectively. Finally, we summarize our technique in Section~\ref{sect-summary}.


\section{Magnetic Field Observations}\label{sect-observations}

Magnetic field observations were obtained through the International Real-time Magnetic Observatory Network \citep[INTERMAGNET]{2013EOSTr..94..373L}.  No baseline was subtracted from the INTERMAGNET data. The observations included one-minute resolution data for the following stations: Brorfelde, Denmark (BFE), Hartland, UK (HAD), Niemegk, Germany (NGK),  Wingst, Germany (WNG), and Alibag, India (ABG). The stations BFE, HAD, NGK, and WNG were selected for being at a similar high magnetic latitude and used as a basis for our UK-specific modeling. We also used  HAD observations of the March 1989 event, obtained from SuperMAG \citep{2009AGUSMSM12A..02G, 2012JGRA..117.9213G}.

ABG data were analyzed to determine the scaling between lower latitudes and higher latitudes. The Alibag Magnetic Observatory is 30 km SSE of Mumbai and was established in 1904 to substitute for the Colaba Observatory, which is the only data source for a non-saturated magnetometer record of the Carrington event. Therefore, comparing the ABG data from recent storms with the higher latitude stations provides an estimate for the proper scaling of the Colaba Carrington event magnetometer measurements to the expected level in the UK, which are unavailable for the Carrington event.  The Colaba data for the Carrington event were 1-hour resolution magnetograms 
\citep{2003JGRA..108.1268T}.

\begin{figure}
\centering

\includegraphics[width=0.85\linewidth]{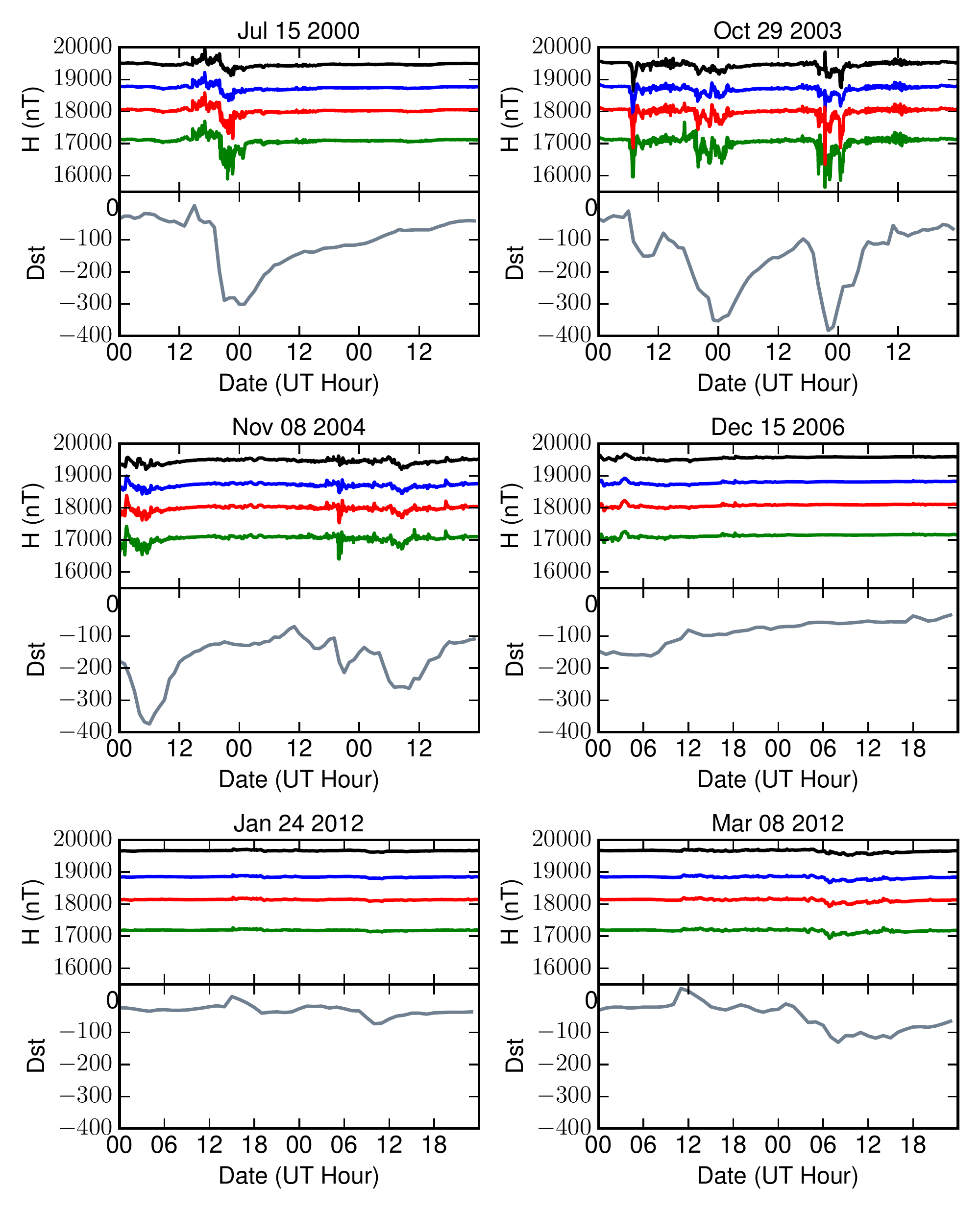}
\caption{Example magnetometer measurements (H is the horizontal component of the magnetic field) for the HAD (black), NGK (blue), BFE (red), and WNG (green) stations, along with the Dst index (gray line) during selected storms. The first three panels starting from the top show intense storms, including the July 2000 Bastille Day event and the Halloween 2003 event. The remaining panels show less intense geomagnetic storms.}
\label{fig-Dst}
\end{figure}

To determine  the time periods for selecting storms, archived observations of the Dst index, which is mainly a global measurement of the ring current \citep{1959JGR....64.2239D} (though contributions from additional currents such as dayside magnetopause current and nightside near-Earth tail current are possible), were used. Using the Dst index provides a consistent metric for geomagnetic activity that is commonly used in space weather analyses. An advantage to using Dst is that it is more highly resolved for large storms than an alternative measurement such as Kp. This is particularly important since we are defining the spectral characteristics, where the rise time of the storm is very important. Dst measurements were obtained from the Data Analysis Center for Geomagnetism and Space Magnetism at Kyoto University.  Strong geomagnetic storms were identified from 1998-2012. Definitions of geomagnetic storms by duration and minimum Dst are somewhat arbitrary, but in general an intense storm is defined as having a minimum Dst $< -100$ and duration $\ge 3$\,hours  \citep{JGRA:JGRA11775}. To select a sample of storms in an automated and uniform way, we adopt the criteria of identifying the most intense storms as those with local minima below -150 for at least 3 hours.  We focused our analysis on the 15 storms following this criteria and with Dst $<$ -200.  Although the Bastille Storm of July 2000 did not meet the time requirement, it was included as one of the most significant storms from the past solar cycle. We also included analysis of weaker storms, including two early 2012 storms associated with solar energetic proton events and two moderate storms from 2001 and 2006. The March 1989 event was included for HAD, since this storm is associated with a large-area power blackout in Quebec. A list of the selected storms is shown in Table~\ref{table-storms}. Figure~\ref{fig-Dst} includes examples of the magnetic field and Dst measurements for a sample of these storms, including magnetometer data from both high intensity and weaker geomagnetic storms.

In all cases, we analyzed the horizontal component of the magnetic field ($B_H$, with units of nT). Where $B_H$ was not included in the data, it was computed as $B_H$ = \begin{math}\sqrt{({B_X}^2 + {B_Y}^2)}\end{math}, where $X$ and $Y$ are the North and East components of the magnetic field. For each storm, $B_H$ is analyzed in the time period 6 hours before the maximum in storm geomagnetic activity, recorded in Table~\ref{table-storms}, until one day later. 

\section{Spectral Scaling}\label{scaling}

In this section, we present our technique for scaling the spectral intensity to reflect a storm of Carrington proportions. The goal of this analysis is to determine the appropriate scaling for the magnetic field from strong geomagnetic storms from recent years to the extreme Carrington event. This comparison is important, since the recent data include higher frequency observations (minute and second scale), while the only available Carrington event data are lower frequency, hour scale observations. The higher frequencies are expected to play an important role in GICs \citep{SWE:SWE20065}, making the use of the more recent magnetometer data particularly crucial. 

\begin{figure}
\centering
\includegraphics[width=0.85\linewidth]{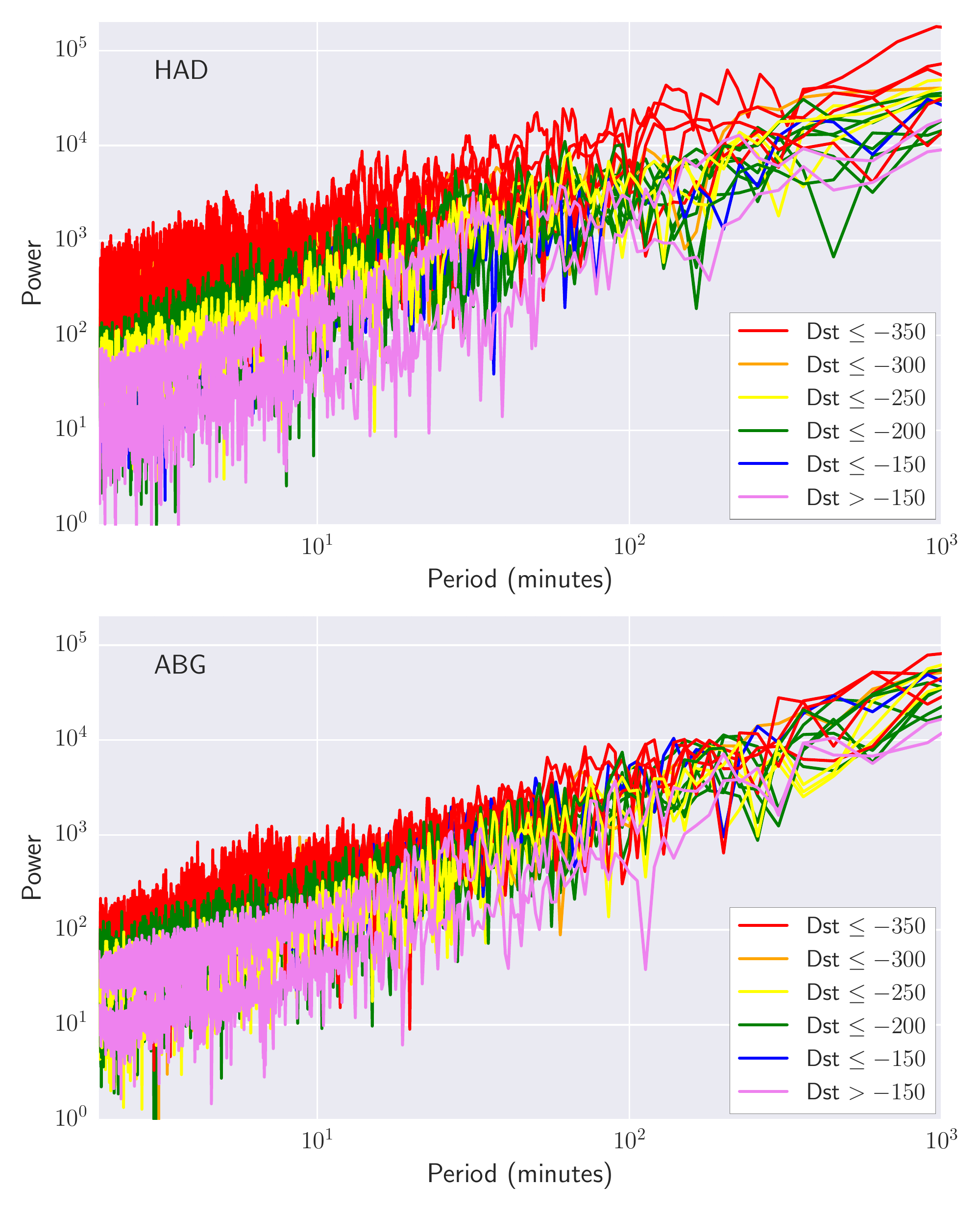}
\caption{The power spectra created from magnetometer observations of all geomagnetic storms in Table~\ref{table-storms} for the HAD (top panel) and ABG (bottom panel) stations. Color-coding is according to the minimum Dst for each storm. The strongest storms are shown in red and the weakest in pink. The slope of the power spectra are approximately the same, while there is a constant scaling factor that is highest for the strongest storms.
}\label{fig-powerspectra}
\end{figure}

The scaling relationships were developed from the power spectra of $B_H$. The purpose of this analysis was to determine the relationship between power spectra from magnetometer measurements and Dst, to derive a relationship that scales spectral components from past storms to create realistic magnetometer measurements of a Carrington-like storm.  Power spectra were created for each geomagnetic storm in Table~\ref{table-storms} for each of the stations (HAD, BFE, NGK, WNG, and ABG). The power spectrum is computed as the absolute value of the discrete fast Fourier transform (FFT). Figure~\ref{fig-powerspectra} shows the power spectra of $B_H$ for all geomagnetic storms analyzed for the HAD and ABG stations. The power spectra of individual storms are color-coded according to the minimum Dst level. As shown in Figure~\ref{fig-powerspectra}, the slope of the power spectra are consistent for all Dst levels, while the stronger storms result in scaled-up power spectra relative to the weaker storm power spectra. We discuss this further and quantify the scaling factors below.

The logarithm of the power spectrum in nT versus the logarithm of the period in minutes (e.g., shown in Figure~\ref{fig-powerspectra} for the HAD and ABG stations) were fit with a linear least-squares regression model as:
\begin{equation}
\log{P} = m_T \log{T} + b_{T}.
\end{equation}
Power ($P$, in units of nT) corresponding to a period from 2 to 90 minutes was used in the fits. We computed the slope ($m_T$), intercept ($b_T$), Pearson correlation coefficient ($r_T$), p-value, and standard error for each storm. Table~\ref{table-scaling} includes the average slope and standard deviation from the 20 geomagnetic storms for each station, as well as the average $r_T$. The linear fits are statistically significant, indicated by $r \approx 0.80$ and low p-values ($< 10^{-6}$ for all fits). The average and standard deviation from the linear fits of the power spectra from HAD are $m_T =1.0919 \pm 0.1296$ and $b_T = 1.4504 \pm 0.3671$. There is no significant variation in the slope with respect to the strength of the storm. However, $b_T$ is dependent on Dst, particularly for the higher latitude magnetometer stations. 

To establish the Dst-dependence of $b_T$, linear least-squares fits were performed using $Dst_{min}$ from Table~\ref{table-storms}. Results of these fits, corresponding to fits of the form
\begin{equation}
b_T = m_{Dst} Dst_{min} + b_{Dst}.
\end{equation}
are included in Table~\ref{table-scaling}. The fits are most significant for the higher geomagnetic latitude stations, with $r_{Dst} \approx -0.8$ and p-values $\le 10^{-5}$. For ABG, the Dst-dependence is less significant, likely due to the lower latitude (i.e., geomagnetic disturbances tend to be larger at high latitudes as an effect of the auroral electrojet \citep{2013SpWea..11..121N}), with $r_{Dst} = -0.55$ and a p-value of 0.019. From our fits, we find a negative correlation between $b_T$ and Dst, resulting in a larger scale factor for stronger storms. 
 
Based on our analysis of the Hartland, UK magnetometer data, using the standard deviation between $m_{Dst}$ and $b_{Dst}$ for the additional stations at similar magnetic latitude, we estimate the expected power for geomagnetic storms in the UK as:
\begin{equation}
	\log P  = (1.0919 \pm 0.1296) \log{T } - (0.00281 \pm 0.00070)\,Dst_{min} + (0.70811 \pm 0.10686).
\end{equation}
This equation provides a means of scaling the frequency components for geomagnetic storms of varying $Dst_{min}$.
For the Carrington event, a wide range of estimates exist for $Dst_{min}$. \citet{JGRA:JGRA16692} estimated $Dst_{min}$ of -1760\,nT, using observations from Colaba. \citet{2006AdSpR..38..173S} reanalyzed the magnetogram with hourly averages, similar to typical calculation of $Dst$, and estimate $Dst_{min} \sim -850$\,nT.

By comparing the power spectrum analysis from the ABG data to the analysis of the HAD data, we estimate the scaling between the Colaba Observatory and HAD. This scaling factor is applied to the Carrington event magnetometer observations from Colaba. For the Carrington event, we find that at a one-hour frequency, the ratio of $P_{UK}/P_{ABG}$ ranges from 4 for $Dst_{min} = -850$\,nT  to as high as 14 for $Dst_{min} =-1760$\,nT. These limits are used as the upper and lower limits for the scaling factor to produce a UK specific Carrington-sized storm. 

\section{Carrington Magnetic Field Model}\label{sect-magfield}
Given these spectral contributions and scaling factors, we construct a magnetic field time series that has characteristics of a Carrington-sized event. The resulting magnetic field model is described as a Carrington-sized storm since we combine scaled data from a selection of storms with varying time resolution. The Colaba data has hourly resolution and forms the basis of the model. We add to this more recent data from smaller-scale geomagnetic storms that have higher cadence time resolution. We also add storm sudden commencement (SSC) separately, based on the March 24, 1991 event, which was one of the fastest moving CMEs in the historical record. Because the SSCs are aperiodic, they can not be scaled using the frequency content of other storms and must be estimated separately. Our magnetic field model, therefore, includes the following: 
\begin{itemize}
\item	one-hour Carrington event data from Colaba, scaled to UK latitudes
\item	one-minute 1989 data from the UK, scaled by $Dst_{min}$ to Carrington-sized storm 
\item	one-second 2012 data from the UK, scaled by $Dst_{min}$ to Carrington-sized storm
\item Storm Sudden Commencement from March 24, 1991 in HAD, scaled by $Dst_{min}$ to Carrington-sized storm
\end{itemize}

The 1989 one-minute data is from the Quebec storm, which is the first storm listed in Table~\ref{table-storms}. This is used for the one-minute resolution data since it is the largest geomagnetic storm in recent years with known GIC effects. Data of one-second resolution have only been available during the current solar cycle, while the intensity of solar activity and geomagnetic storms has been less than in the previous solar cycles. Therefore, we must choose a weaker storm for the one-second time frequency observations. The March 2012 storm (storm 20 in Table~\ref{table-storms}) is among the strongest with reliable one-second data.   

The scaling factors were derived used the results from the power analysis in Section~\ref{scaling}. The one-hour scaling factor is the ratio of $P_{UK}/P_{ABG}$ at $Dst_{min} =-1000$\,nT, with limits described in Section~\ref{scaling}. The one-minute scaling factor is computed using Equation~3 as $P_{(Dst = -1000)}/P_{(Dst = -589)} \sim 15$, from the ratio of the computed power of the Carrington storm to the March 1989 storm. Since the one-second data are not available for a wide range of storm strengths, we could not use the same technique to reliably determine the highest frequency scaling component. Instead, we assume that the highest frequency component is consistent for weaker storms, with  $|Dst_{min}| < 250$\,nT, and scale the 2012 one-second data by the same factor of 15. Though, our scaling relation in Equation~3 suggests the factor could be as high as ten times this amount. More stringent limits on the one-second scaling will require observations of higher intensity storms, of which HAD observations are only available for storms with $|Dst_{min}| < 200$\,nT. For the SSC component, we scaled the March 24, 1991 storm's SSC by a factor two, chosen to match the level of the positive peak in the one-hour Carrington event data from Colaba. 

\begin{figure}
\centering
\includegraphics[width=0.85\linewidth]{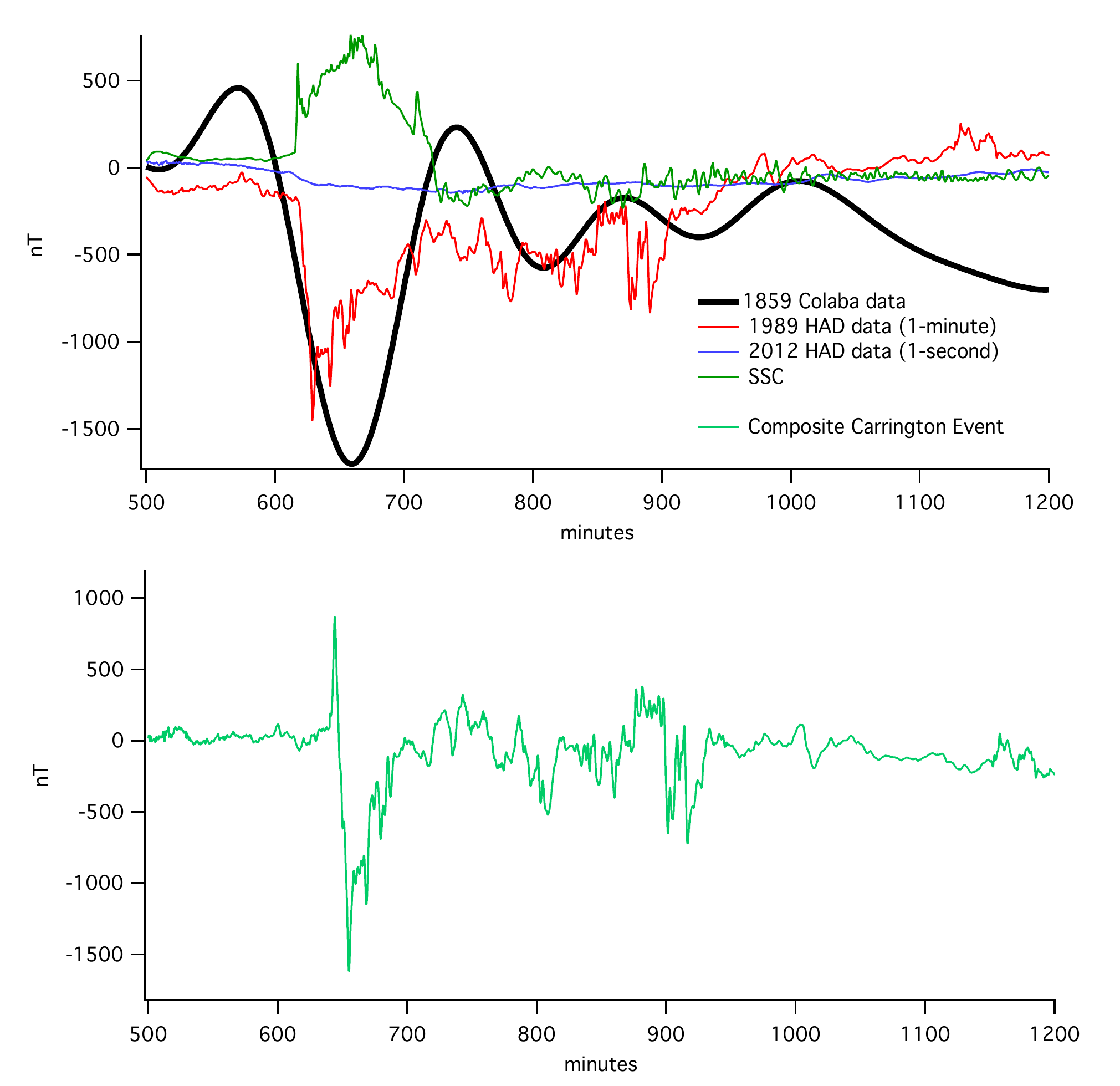}
\caption{Magnetic field profile derived from Colaba and modern HAD observations. The top panel shows the individual contributing storm components (full details on the scaling factors are described in Section~\ref{sect-magfield}). The black line shows the 1-hour resolution magnetometer observation from the Colaba observatory, scaled to the geomagnetic location of the UK. The scaled HAD magnetometer observations are shown for the 1-minute (red) and 1-second (blue) components of B$_H$ from the March 1989 event and a March 2012 storm, respectively. The scaled storm sudden commencement component from the HAD observations of the March 24, 1991 storm is shown in green. The bottom panel shows the composite magnetic field event, created from the above mentioned components.
}\label{fig-magfieldmodel}
\end{figure}

We extract each frequency band from the listed events, resample each to a one-second time resolution, and add them in the time domain. Figure~\ref{fig-magfieldmodel} shows this composite time series. For the composite time series, a linear detrend was implemented to ensure that there is no step between the first and last data points. Additionally, the direct component offset of the magnetic field was zeroed, since for the electric field calculation this component will not contribute to E, which is a function of dB/dt.
This magnetic field model will be used in the following sections as the magnetic field driver to a GIC impact analysis of Carrington-type event on a modern UK power system. 

The frequency bands were extracted using wavelet coefficients in a multi-resolution analysis. While similar results are achieved with a FFT with a Gaussian window, we chose a wavelet analysis because it is simpler to deal with aperiodic components and for previous use of the technique in magnetometer studies (e.g., \citealt{2008JASTP..70.1579X} and \citealt{Survey:2013fk}). The output of the wavelet transform are 12 levels of frequency-banded time series. For example, there were 3 time series with characteristic frequencies with in the 1-minute to 1-hour band. These were selected, and recombined using a wavelet transformation in the reverse direction. This is similar to a band-bass filter in frequency space analysis, but preserves the aperiodic fluctuations that are important to GIC calculations.

\begin{figure}
\centering
\includegraphics[width=0.85\linewidth]{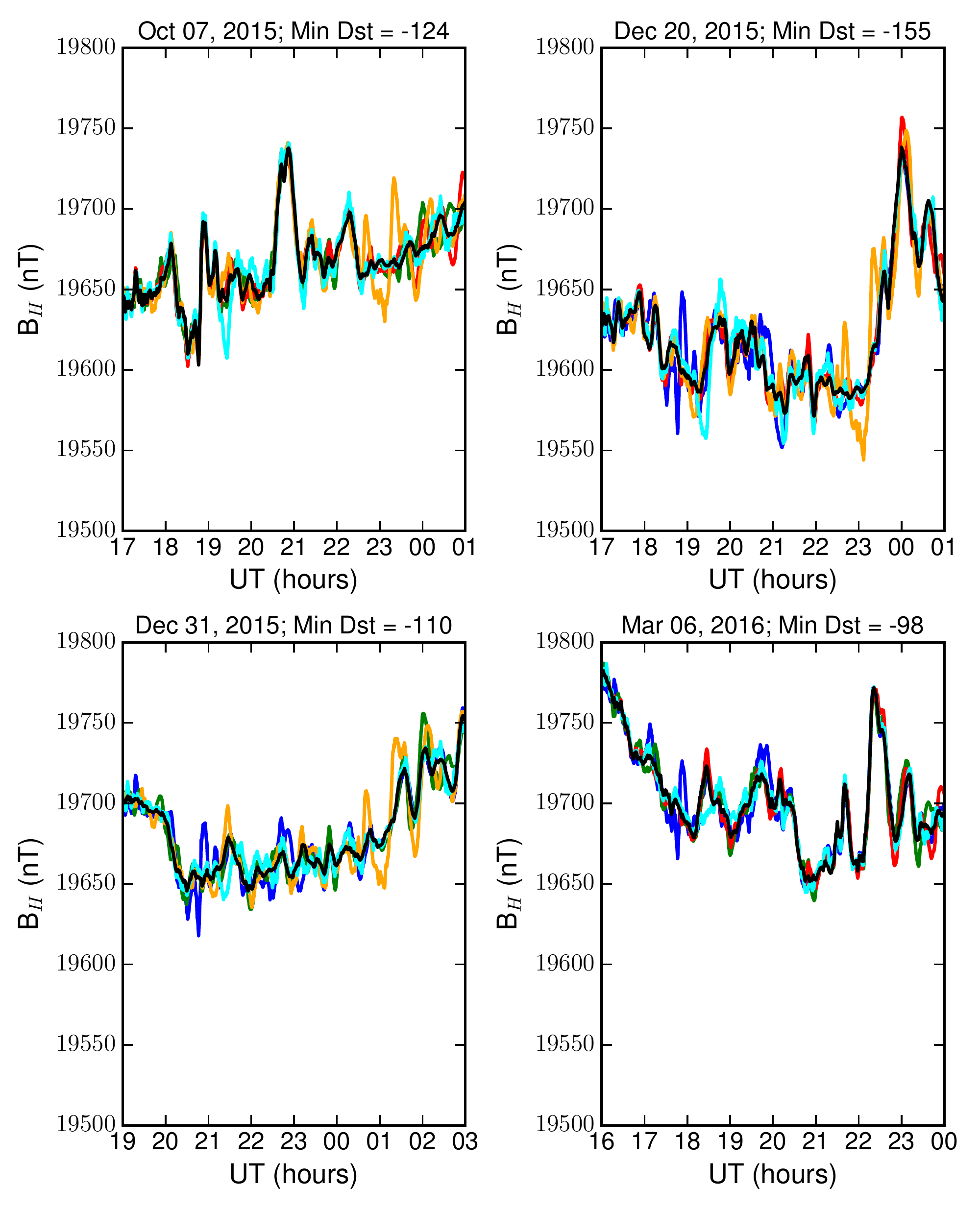}
\caption{Simulated magnetometer observations for recent geomagnetic storms. The HAD magnetometer observations for the specified storm are shown in black. Simulated data were created by combining the hourly component of the observation with scaled minute and second resolution components from each of the indicated storms. The colors of the simulated storms correspond to: blue as Dec. 20, 2015; cyan as Oct. 7, 2015; green as Jan. 1, 2016; red as Mar. 6, 2016; and orange as May 8, 2016. The average normalized rms error between the simulated and observed data for these similar-sized storms is 5-8\%.
}\label{fig-simulatedseconds}
\end{figure}

To test our magnetic field scaling method, we applied the technique to HAD second-resolution magnetometer observations of five recent geomagnetic storms with minimum Dst levels ranging from -83 to -155 nT. These storms are similar intensity to the 2012 data used for the Carrington-sized storm scaling. As a first step, we performed the same power spectrum analysis on the second-resolution data as the minute-resolution data analysis described in Section~\ref{scaling}. We found that the higher frequency data scaled in a manner consistent with the results of the minute-resolution analysis. We then used our technique to simulate the magnetometer observations at HAD for each of the storms by using the one-hour resolution component of the individual storm along with scaled minute and second resolution components from each of the additional four storms. Results of the analysis are shown in Figure~\ref{fig-simulatedseconds}. We find that our scaling technique produces realistic magnetic field models. The average normalized rms error computed between the magnetic field measurements and the simulated models using the scaled frequency components of different storms ranges from 5-8\% with a standard deviation of 0.015. The error introduced from using higher frequency components from different storms is therefore much lower than the range of Dst values estimated for the Carrington event (between -850 and -1760 nT).

\section{Conductivity Models}\label{sect-conductivity}
Given the magnetic field constructed for the model event, we next determine how that magnetic field interacts with the conductivity of the Earth beneath the power systems in our analysis. There are several models of deep earth conductivity available through the European Risk from Geomagnetically Induced Currents (EURISGIC) program
\citep{2011SpWea...9.7007V}. These are simple, 1-D layer cake models with three or four layers per region as represented in Table~\ref{table-conductivity}. 

While the magnetic field is the driver of GIC, deep Earth conductivity determines how large of a response the Earth will produce in terms of an induced electric field. A complete analysis of electric field induction would require a detailed 3-D conductivity model of the UK.  The EURISGIC models used in this example are simplified, and are intended to provide a first order estimate. As such, they do not include 3-D structure, an omission which can lead to electric field reductions along boundaries.  In addition, the EURISGIC models are not very well resolved at shallow depths. This means that electric field induction due to the higher frequency magnetic field fluctuations will be underestimated in our analysis. 

\citet{SWE:SWE20065} performed an analysis including 3-D conductivity models with detailed shallow conductivity. They found that while detailed conductivity makes a difference in the calculation, it is a second order effect after the magnetic field driver. In general, better conductivity models lead to increased electric field estimates. This again implies that our estimations using 1-D conductivity may be too low and should be used as a lower bound. 

The coastal effect is a good example of electric field enhancements that can occur due to lateral variations in conductivity. At a water-land interface, there is a very sharp difference in conductivity. This phenomenon is observed in \citet{2002JASTP..64.1779B} and \citet{2002JASTP..64..743E} for the UK system and modeled by 
\citet{2004EP&S...56..525O} and \citet{2005SpWea...3.4A03G}.  The models demonstrate the importance of including the sea conductivities, as the magnetic field perturbations deduced with a 3-D conductivity model in \citet{2004EP&S...56..525O}, which include the sea conductivity, are noticeably closer to the magnetometer observations compared to either the 1-D conductivity model (which omit sea water conductivity) or the Dst-based model. \citet{2005SpWea...3.4A03G} produced a different model and presented figures on the resulting voltage enhancements due to separate, Òreasonable,Ó worst-case sudden storm commencement and electrojet scenarios.  The enhancements were due to the presence of the ocean (which produced abrupt changes in conductivity) and approached 250 V in the sudden storm commencement scenario and 400 V in the electrojet case (assuming a Gaussian rather than uniform magnetic field).  

In the applications to the UK power grid, the risk analysis in \citet{2002JASTP..64..743E} showed that transformers at the coasts exhibited the highest GIC flows and associated saturation.  Quantitatively, \citet{2002JASTP..64.1779B} found that the contrast in resistivity at the coastal boarders produced electric field enhancements as high as 4 V km$^{-1}$ for an auxiliary magnetic field (H = 1 A m$^{-1}$ or $B \approx 1256$\,nT) and a period of 10 minutes, a perturbation that approaches the electric field fluctuations measured by  \citet{2005SpWea...311002T} during the Halloween storm in 2003. This is consistent with 100-year values at HAD 
reported by \citet{2011SpWea...910001T} and estimates for similar latitudes in \citet{2016SpWea..14..668W}.


\section{Induced Electric Field}\label{sect-efield}
To calculate the electric field with 1-D conductivity profiles, we first calculate the surface impedance as a function of frequency for each of the four UK conductivity models. We use the standard plane wave assumption and the method described in \citet{weaver1994mathematical} to produce the surface impedance profiles. 
There are small differences in the impedance profiles for each of the 4 models, but they are generally similar.

\begin{figure}
\centering
\includegraphics[width=0.95\linewidth]{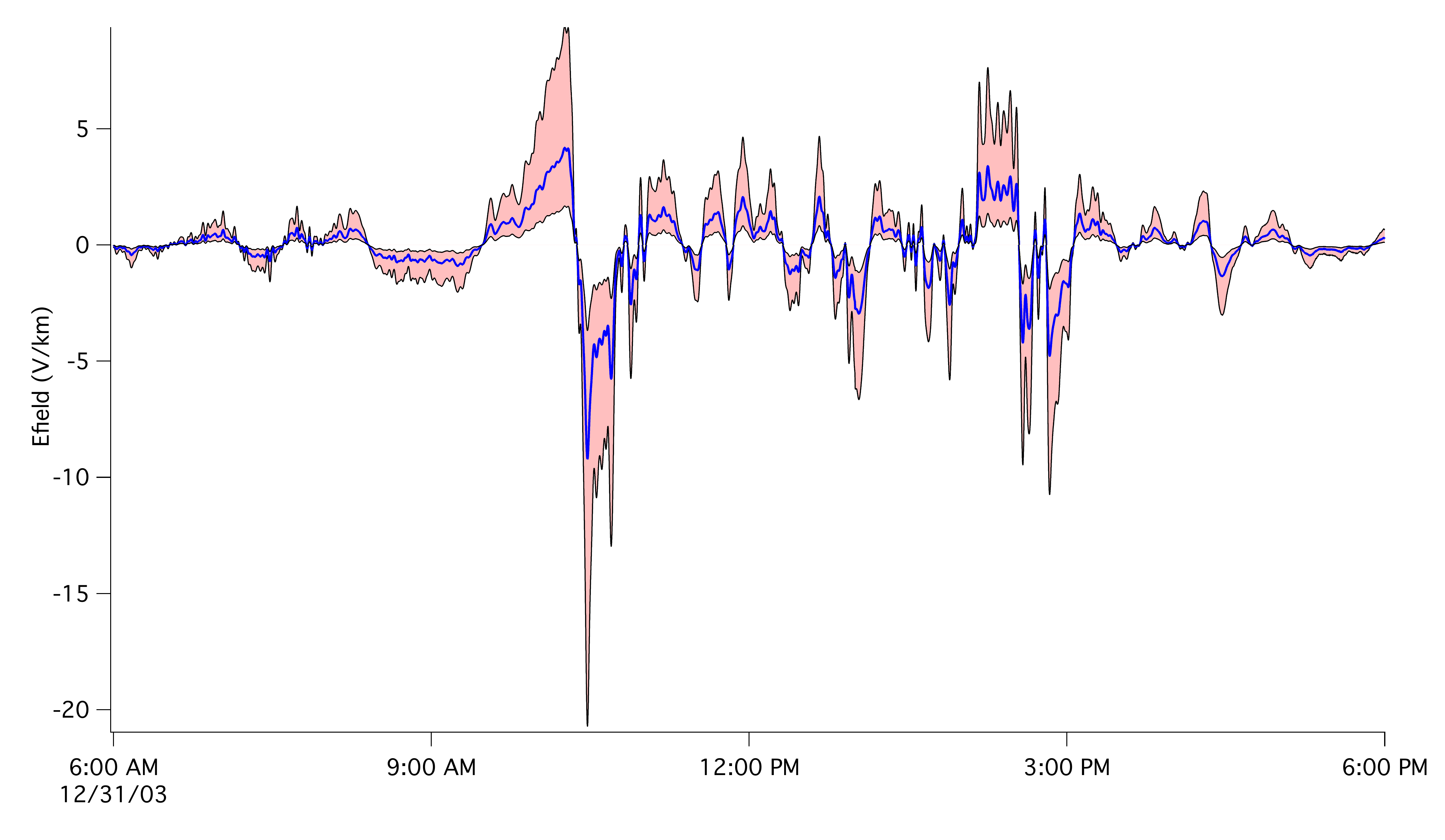}
\caption{Electric field model and estimated uncertainty (red shaded region) induced by a Carrington-level event (blue curve).  The range in uncertainty is determined by the uncertainties in fitted parameters given in Table~\ref{table-scaling}.
}\label{fig-efield}
\end{figure}

The electric field at a given spectral component is calculated in the frequency domain using a method conceptually similar to \citet{2001JGR...10621039T}, which relates the electric and magnetic fields in the following way:
\begin{equation}
	E(\omega)= B(\omega)Z(\omega) ,	
\end{equation}
where E($\omega$) is the electric field, B($\omega$) is the magnetic field, and Z is the surface impedance at frequency $\omega$. Using the surface impedance profiles from Table~\ref{table-conductivity},
and applying a wavelet method to transform between the frequency and time domains, we produce the electric field time series shown in Figure~\ref{fig-efield}.

\begin{figure}
\centering
\includegraphics[width=0.85\linewidth]{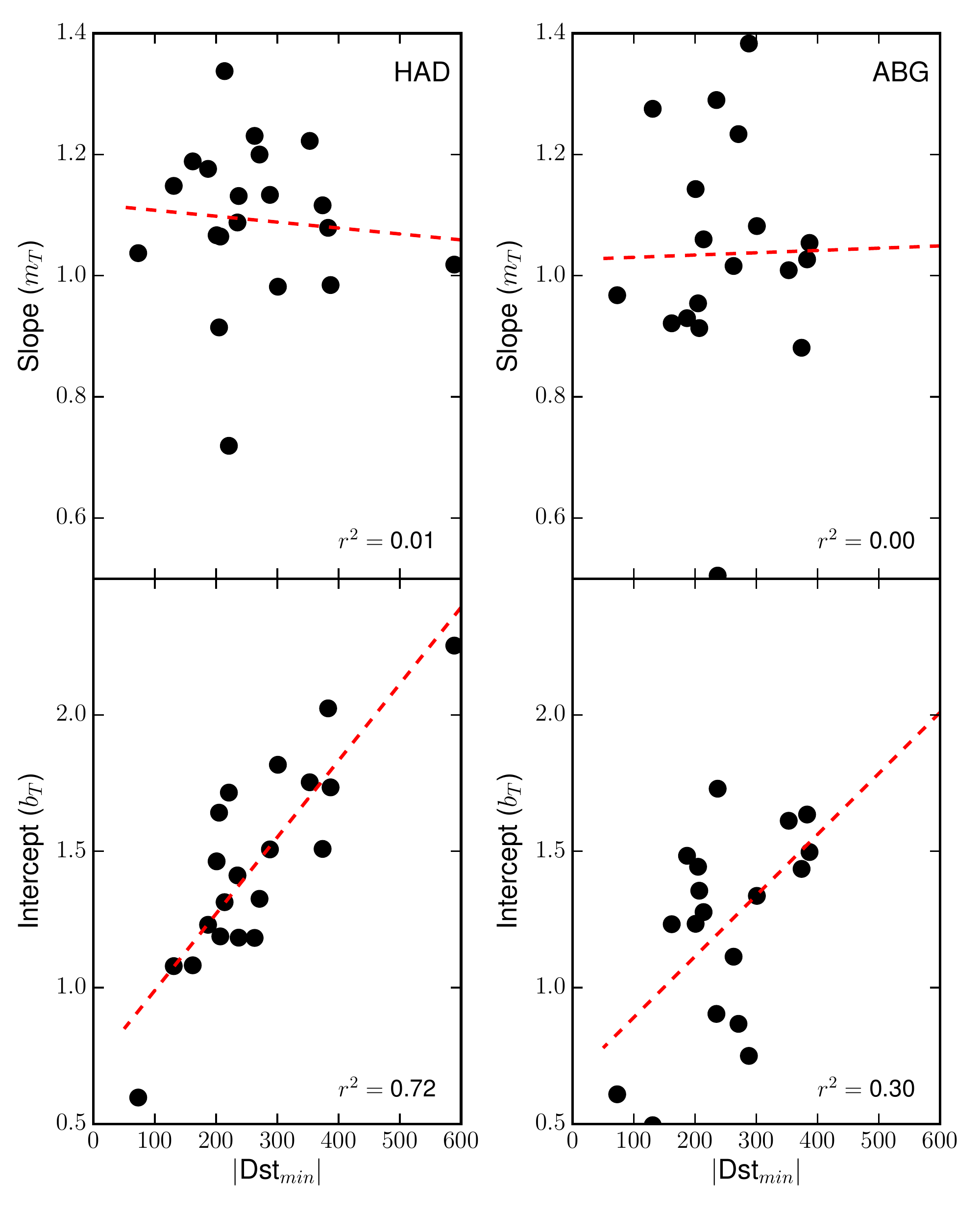}
\caption{Examples of linear least-squares fits to determine the Dst-dependence of the power spectra-time best-fit parameters for the HAD and ABG stations. The slope ($m_T$) and intercept ($b_T$) are derived from fits to the power spectra of each of the geomagnetic storms listed in Table~\ref{table-storms}. No Dst-dependence is found for $m_T$, with a Pearson correlation coefficient $< 0.1$ and p-value $\approx 1$ for all stations. A correlation exists with $b_T$, which we quantify in Table~\ref{table-scaling}.
}\label{fig-powerspectrafits}
\end{figure}

In Section~\ref{scaling}, we used a scaling technique to adjust the spectral content between storms of different intensity. However, every storm is unique, and the relationship between spectral enhancements is not exact. This uncertainty can be seen in the scatter of values around the line fits in Figure~\ref{fig-powerspectrafits} and is represented as a range of values given in Table~\ref{table-scaling} and Equation 3. Using the largest and smallest combinations of scaling parameters, we estimate the range of electric field that is possible within these bounds. These bounds are shown in Figure~\ref{fig-efield} as the red range surrounding the blue line, which is our electric field estimate. Given this uncertainty propagation, we estimate that the range of electric field values at the maximum point is approximately 4-20 V km$^{-1}$. This is a wide range of values that takes into account the possibility of a unique storm that does not scale in a typical or average manner. 

The scaling uncertainties derived from Table~\ref{table-scaling} are a result of the lack of direct measurements of extreme storms in the historical record. There are other sources of error in our simple analysis that are due to incomplete information. We have mentioned these uncertainties in previous sections, but reiterate them here to support the idea that these electric field estimates and GIC estimates should be used as a lower bound on the potential impact of a Carrington event. Each of the following has the potential to impact the electric field estimates for a given storm:
\begin{enumerate}
\item	Lack of shallow structure and lateral variations in conductivity models. This analysis uses EURISGIC 1-D conductivity models. These are very simple layer cake models that do not adequately (for analyses of this type) specify the complicated 3-D structure of conductivity variations.
\item	Incomplete one-second resolution magnetic field data. This analysis uses a very small event from 2012 to determine the spectral scaling of high frequency content. It is likely that extreme events have a completely different high frequency spectrum. We have more confidence in the scalability of the one-minute magnetic field data because of the large events that have been recorded at this time resolution.
\item	Lack of coast effects. It has been well-established that induced electric fields are enhanced along the coasts, with the potential effect extending inland 70-100 km. The range of estimates of this impact varies considerably, but it is likely that the electric field could be significantly enhanced due to the coastal effect.
\end{enumerate}
By taking these sources of uncertainty into consideration in future analysis, the errors in the impact of an extreme event can be reduced. Better conductivity models, as well as magnetic field records with a one-second time resolution for other moderate events, are available through BGS. 

With these uncertainties in mind, our electric field calculated for the Carrington-type event has a maximum intensity of approximately 9 V km$^{-1}$. In comparison, the proposed NERC guidelines for GIC mitigation in the US, TPL-007-1 , include a maximum electric field magnitude of 8 V km$^{-1}$ for a 1-in-100 year reference event. \citet{2008SpWea...6.7001P} estimate a Carrington-level electric field of 4 V km$^{-1}$ based on solar wind correlations. In contrast, \citet{2003SpWea...1.1016K} suggests that a 1-in-100 year event should be closer to 20 V km$^{-1}$, and the Carrington event could have been even larger. \citet{SWE:SWE20065} reported that 10 V km$^{-1}$ would be expected in the UK for an extreme storm, based on 3-D conductivity and modeling results, although values of near 30 V km$^{-1}$ were possible in their model. The variation in these numbers is a reflection of the uncertainties in many aspects of these analyses. We suggest that, while our simple analysis is consistent with the more advanced analysis, it be used as a lower bound on potential impacts due to the uncertainties already discussed. The advantage of our technique over alternative methods is that frequency-domain analyses are simple to compute, while still preserving aperiodic signals (e.g., the storm sudden commencement). During large storms, the sudden storm commencement is completely aperiodic, and it can contribute very strongly to the induced geo-electric field.

\section{Summary}\label{sect-summary}

Even for modern large storms there is a high level of uncertainty in the actual intensity and effect of GICs on power grid systems. The higher time resolution observations of the magnetic field in the past decade have not recorded any large storms. Conductivity models are simplified and the possible impacts of local variations can be large. Further, the actual impact and susceptibility of transformer assets to an induced GIC is highly dependent on transformer type and configuration.

Despite these caveats, we present a technique to model a Carrington-like event, which gives an estimate of the direct impacts of an extreme event based on available data and models. The actual impacts during a real storm will vary with the specifics of a storm and within the uncertainty of the assumptions we made. As an example, we determined the relationship between storm strength, parameterized with $Dst$ and the horizontal component of the magnetic field for five magnetometer stations using twenty geomagnetic storms. Using the scaling relations, we combined magnetic field measurements from representative storms to create a model magnetic field for the UK. Using a ground conductivity model, we determined an extreme event electric field model.

Based on our extreme event scenario and a comparison to published work (e.g., \citet{SWE:SWE20065}), electric field values of 10 V km$^{-1}$ could reasonably be expected for the UK during an extreme event. GICs induced during these events could be impactful, in the range of 100-300 A, depending on asset location and system orientation specifics. The impact on power grid assets would be highly dependent on transformer configuration, system topology, and the spectral content of a given storm.

\acknowledgements
The results presented in this paper rely on data collected at magnetic observatories.  We thank the national institutes that support them and INTERMAGNET for promoting high standards of magnetic observatory practice (data publicly available at: \url{http://www.intermagnet.org}). We also gratefully acknowledge SuperMAG, PI Jesper W. Gjerloev. The Dst index data were obtained from the World Data Center for Geomagnetism, Kyoto, available at \url{http://swdc234.kugi.kyoto-u.ac.jp/index.html}. The data necessary to reproduce the analysis are available from the authors upon request ({lisa@astro.io}).

\newpage

\end{article}

\clearpage

\clearpage
\begin{table}[h]
\begin{center}
\caption{List of Geomagnetic Storms Included in Scaling Analysis}\label{table-storms}
\begin{tabular}{c c c c}
\hline\hline
$N$ & Storm Date & $T_{Dst_{min}}$ & $Dst_{min}$ \\
\hline
1 & 03/13/1989 & 0 & -589\\
2 & 05/04/1998 & 5 & -205 \\
3 & 09/25/1998 & 9 & -207 \\
4 & 10/22/1999 & 6 & -237 \\
5 & 04/07/2000 & 0 & -288 \\
6 & 07/16/2000 & 0 & -301 \\
7 & 08/12/2000 & 9 & -235 \\
8 & 09/17/2000 & 23 & -201 \\
9 & 03/31/2001 & 8 & -387 \\
10 & 04/11/2001 & 23 & -271 \\
11 & 10/21/2001 & 21 & -187 \\
12 & 11/24/2001 & 16 & -221 \\
13 & 10/30/2003 & 0 & -353 \\
14 & 10/30/2003 & 22 & -383 \\
15 & 11/08/2004 & 6 & -374 \\
16 & 11/09/2004 & 21 & -214 \\
17 & 11/10/2004 & 10 & -263 \\
18 & 12/15/2006 & 7 & -162 \\
19 & 01/25/2012 & 10 & -73 \\
20 & 03/09/2012 & 8 & -131 \\
\hline \\
\end{tabular}
\end{center}

Details for the twenty selected storms used in determining the scaling factor between Dst and the magnetometer observations. Included are the storm number ($N$), date of the onset of the storm (Storm Date), minimum Dst time in UT hours ($T_{Dst_{min}}$), and minimum measured Dst ($Dst_{min}$).

\end{table}

\begin{table}[h]
\begin{center}
\caption{Results from the Scaling Analysis of the Power Spectra}\label{table-scaling}
\begin{tabular}{c c c c c c c c c}
\hline\hline
Station & $\lambda_B$ ($^{\circ}$ N) & $\phi_B$ ($^{\circ}$ E) & $m_T$ & $m_{Dst}$ & $b_{Dst}$ & $r_T$ & $r_{Dst}$ & p-value \\
\hline
HAD & 53.64  & 80.28  &   $1.0919 \pm 0.1296$ & -0.00281 & 0.70811 & 0.81 & -0.85 & $3 \times 10^{-6}$\\
BFE & 52.05  & 89.11  &   $1.0546 \pm 0.3894$ & -0.00432 & 0.60264 & 0.82 & -0.83 & $7 \times 10^{-5}$\\
NGK & 51.84 & 97.71  &   $1.0885 \pm 0.4029$ & -0.00318 & 0.51917 & 0.75 & -0.86 & $8 \times 10^{-6}$\\
WNG & 54.06 & 95.04  &    $1.2039 \pm 0.1610$ & -0.00440 & 0.41762 & 0.84 & -0.86 & $1 \times 10^{-5}$\\
ABG & 10.37  & 146.55  &   $1.0359 \pm 0.1906$ & -0.00224 & 0.66671 & 0.82 & -0.55 & $2 \times 10^{-2}$\\
\hline \\
\end{tabular}
\end{center}

Results from linear least-squares fits to the power spectra from the twenty storms listed in Table~\ref{table-storms}. The geomagnetic latitude ($\lambda_B$) and longitude ($\phi_B$) are shown in degrees for each station. Separate fit parameters are given for each of the magnetometer stations for which data were analyzed. A preliminary fit to the logarithm of the power spectrum as a function of logarithm of time ($T$, minutes) gives the slope ($m_T$) and the Pearson correlation coefficient ($r_T$). A second fit to the intercept of this first fit as a function of Dst yield the slope ($m_{Dst}$) and intercept ($b_{Dst}$), along with $r_{Dst}$ and the p-value. This gives the power as a function of Dst and time: \begin{math} \log{P} = m_T \log{T} + m_{Dst} Dst_{min} + b_{Dst}\end{math}.

\end{table}

\begin{table}[]
\centering
\caption{EURISGIC 1-D Conductivity Models}\label{table-conductivity}
\begin{tabular}{ ll | ll | ll | ll }
\hline\hline
 M20 &  & M21  &  & M22  &  & M23  &  \\
 \hline\hline 
 d (km) & $\rho$ ($\Omega$m)  & d (km)  & $\rho$ ($\Omega$m)  & d (km)  & $\rho$ ($\Omega$m)  & d (km)  & $\rho$ ($\Omega$m)  \\
\hline
 20 & 1000  &  10 & 100  & 10  & 500  & 20  & 1000  \\
 $\infty$ &  200 &  20 & 150  & 20  & 2000  & 30  & 5000  \\
 &  &  30 & 250  & 30  & 500 & $\infty$  & 200  \\
 &  &  $\infty$ & 200 & $\infty$ & 200 \\
 \hline 
\end{tabular}

Values from the 1-D Geoelectric Litosphere Model of the Continental Europe available from EURISGIC. We use models M20, M21, M22, and M23. The values of conductivity, $\rho$ ($\Omega$m), for a given distance, d (km), are included here.
\end{table}

\end{document}